\documentclass[aps,epsf,
superscriptaddress,
twocolumn,prl]{revtex4}
\newcommand{\expec}[1]{\left < #1\right >}

\usepackage{amsmath}
\usepackage{amsfonts}
\usepackage{graphicx}
\usepackage{amsthm}
\usepackage{hyperref}
\usepackage[usenames]{color}
\renewcommand{\v}[1]{\mathbf{#1}}

\newcommand{\lp}{\left ( }
\newcommand{\rp}{\right ) }
\newcommand{\lb}{\left [ }
\newcommand{\rb}{\right ] }

\newcommand{\beq}{\begin{eqnarray*}}
\newcommand{\eeq}{\end{eqnarray*}}

\newcommand{\be}{\begin{eqnarray}}
\newcommand{\ee}{\end{eqnarray}}

\newcommand{\mc}{\mathcal}

\def\lsim{\mathrel{\rlap{\lower4pt\hbox{\hskip1pt$\sim$}}
    \raise1pt\hbox{$<$}}}                

\def\gsim{\mathrel{\rlap{\lower4pt\hbox{\hskip1pt$\sim$}}
    \raise1pt\hbox{$>$}}}                

\usepackage{subfiles}

\begin{document}

\title{Universality class of quantum criticality in the two-dimensional Hubbard model at intermediate temperatures ($t^2/U\ll T\ll t$) }
\author{Kaden R.~A. Hazzard} \email{kaden.hazzard@colorado.edu}
\affiliation{JILA and Department of Physics, University of Colorado, Boulder, and NIST, Boulder, Colorado 80309-0440, USA}
\author{Ana Maria Rey}
\affiliation{JILA and Department of Physics, University of Colorado, Boulder, and NIST, Boulder, Colorado 80309-0440, USA}
\author{Richard T. Scalettar}
\affiliation{Physics Department, University of California, Davis, California 95616, USA}

\begin{abstract}

We show that the dilute Fermi gas quantum critical universality class quantitatively describes the Mott/metal crossover of the two-dimensional Hubbard model for temperatures somewhat less than (roughly half) the tunneling but much greater than  (roughly twice) the superexchange energy.  We calculate the observables expected to be universal near the transition --- density and compressibility --- with numerically exact determinantal quantum Monte Carlo.  We find they are universal functions of the chemical potential. Despite arising from the strongly correlated regime of the Hubbard model, these functions are given by the weakly interacting, dilute Fermi gas model.  These observables and their derivatives are the only expected universal static observables of this universality class, which we also confirm by verifying there is no scaling collapse of the kinetic energy, fraction of doubly occupied sites, and nearest neighbor spin correlations. Our work resolves the universality class of the intermediate temperature Mott/metal crossover, which had alternatively been proposed to be described by more exotic theories.  However, in the presence of a Zeeman magnetic field, we find that interplay of spin with itinerant charge can lead to physics beyond the dilute Fermi gas universality class.
\end{abstract}

\maketitle


\textit{Introduction.}---The crossover between a Mott insulator (MI) and a metal in strongly correlated materials is an incredibly rich problem~\cite{imada_metal-insulator_1998}.  At least two key factors play a role: the emergence of a charge gap and the onset of magnetic order due to superexchange, and the role of each was debated originally by Slater and Mott in the 1940's and 50's.  These intertwined effects --- the former of which is in general indescribable by a local order parameter --- give small energy scales at the transition and excitations proliferate, and consequently impede a full theoretical treatment.  A quantum critical perspective has frequently been applied to the two-dimensional Hubbard model, the simplest canonical model that includes MI and metallic states.  At temperatures low compared to the superexchange temperature there are numerous proposed scenarios but few definitive conclusions, and, more basically, the \textit{phases} at zero temperature are unknown.  Moreover, open questions remain for even the simplest quantum critical scenarios (e.g., \cite{sachdev_where_2010}).

We address the related question at temperatures \textit{above} the superexchange scale, showing that in an appropriate regime the ``dilute Fermi gas" (DFG) universality class quantitatively describes the low energy behavior.
Although the physics at these temperatures  is simpler than at low temperatures, it has remained uncertain whether and in what regime universal physics manifests, and even less clear which universality class describes the behavior\cite{imada_metal-insulator_1998}.
 Although a common scenario of the doping-driven MI/metal crossover at these temperatures involves the DFG theory where the effective number of carriers vanishes at the (avoided or fictional) zero-temperature transition, such behavior has not been verified with exact calculations~\cite{imada_metal-insulator_1998}.  Other possibilities exist, for example a diverging effective mass at the transition. Another possibility comes from zero-temperature determinantal quantum Monte Carlo, which indicates a dynamic critical exponent $z=4$, in contrast to $z=2$ predicted by the DFG and other low temperature theories~\cite{imada_metal-insulator_1998}. This has been conjectured to persist up to temperatures above the superexchange scale~\cite{imada_metal-insulator_1998}.
 We show through analysis of numerically exact  determinantal quantum Monte Carlo calculations that the DFG provides a quantitative description in an appropriate regime.

We consider the two-dimensional square lattice Hubbard model defined by the Hamiltonian
\be
\hspace{-0.17in}H \!\!&=& \!\! -t \!\!\sum_{ \langle i,j\rangle,\sigma} \!\! f^\dagger_{i\sigma} f_{j\sigma}^{\phantom{\dagger}} + \sum_i \lb (U/2)
  n_i(n_i\!\!-1)
-\mu n_{i} + h s_{iz}
\rb\label{eq:Hubb}
\ee
with $f_{i\sigma}^{\phantom{\dagger}}$ and $f_{i\sigma}^\dagger$ fermionic annihilation and creation operators for states at site $i$ and spin state $\sigma\in \{\uparrow,\downarrow\}$,
$n_{i\sigma}\equiv f^\dagger_{i\sigma}f_{i\sigma}^{\phantom{\dagger}}$, $n_i \equiv n_{i\uparrow}+n_{i\downarrow}$, $s_{iz}\equiv n_{i\uparrow}-n_{i\downarrow}$, the sum $\sum_{\langle i,j\rangle}$ indicating a sum over nearest neighbor sites $i$ and $j$, $t$ the tunneling, $U$ the interaction energy, $\mu$ the chemical potential, and $h$ an external Zeeman field.  We defined the chemical potential  such that the center of the half-filling $n_{i}=1$ MI at $t=0$ occurs at $\mu=0$.

We  use
determinantal quantum Monte Carlo (DQMC)~\cite{blankenbecler_monte_1981} to compute spin densities $n_\sigma=\expec{n_{\sigma,i}}$, compressibility $\kappa=\partial n/\partial \mu$, kinetic energy $K = -t\sum_{\langle j\rangle,\sigma} \langle f^\dagger_{i\sigma}f_{j\sigma}^{\phantom{\dagger}}\rangle$ (sum over neighbors $j$ of arbitrary site $i$), doubly occupied site fraction $D=\expec{n_{i\uparrow}n_{i\downarrow}} $, and the nearest-neighbor spin correlator $X=\expec{s_{iz}s_{jz}}$ with $i$ and $j$ nearest neighbors.
 We show that for $h=0$, small enough $t/U$, and $T$ in the window $t^2/U \lsim T\lsim t$ (we set $k_B=1$), the DFG universality class quantitatively describes the $\mu$-dependence of the expected universal observables ($n$ and $\kappa$) near the MI/metal crossover.  We also confirm that, as predicted by the DFG theory, other observables are non-universal.  In contrast, we find that the Zeeman field dependence of the observables is not well described by the DFG or any $z=2$ universality class, and perhaps is described by another, unknown universal theory.

Many of these observables can be measured in cold fermionic atoms trapped in optical lattices~\cite{joerdens_mott_2008},
which are already in the $T\sim t$ regime. Thus, one can apply a similar analysis to that carried out here.  In the present region, one can use the comparison to validate the faithful emulation of the Hubbard Hamiltonian.  As these experiments achieve somewhat colder temperatures, they will be able to explore more exotic quantum criticality  inaccessible to DQMC~\cite{zhou_signature_2010}.
Even at the present temperatures, these may yield insight into quantum critical spectra and dynamics that are unobtainable numerically.


\textit{Scaling.}---For conventional quantum phase transitions described by symmetry breaking of a bosonic order parameter,  observables $\mc O$ take the form~\cite{sachdev_quantum_2001}
$
\mc O(g,T,\ldots)
=
\mc O_r(g,T,\ldots)
+ T^{1+d/z-1/(z\nu_{\mc O})} \Psi_{\mc O} \lp (g-g_c)/T^{1/(z\nu_g)}\rp \label{eq:scaling-gen}
$
where $g$ is a relevant coupling, the ``$\ldots$" of the argument indicate irrelevant couplings, $\nu_{\mc O}$ and $\nu_g$ are, possibly non-trivial, scaling dimensions, $d$ is the spatial dimension, and $z$ is the dynamic critical exponent.  The function $\mc O_r$ is analytic and $\Psi_{\mc O}$ is a singular, universal contribution.  The definition of ``universality" is that the only dependence of $\Psi_{\mc O}$ on the irrelevant variables is to change the dimensionful constant in $\Psi_{\mc O}$ that converts $\frac{g-g_c}{T^{1/(z\nu_{g})}}$ into a dimensionless constant. These constants for $z=1$ and $z=2$ theories have the interpretation of an effective sound speed and mass for the excitations, respectively. We have assumed $g$ is the only relevant coupling.  Generally there can be more than one relevant variable, but frequently there are only one or two.

For conserved observables $\mc O$ such as density and magnetization and their dependence on variables coupling to these quantities, the  scaling structure simplifies to~\cite{sachdev_quantum_1994}
\be
\hspace{-0.17in}{\mc O}(\mu,h, T\!,\!\ldots)\!\!&=&\!\! {\mc O}_r(\mu,h,T\!,\!\ldots)
    + T^{x} \Psi_{\mc O}\!\!\lp \! \frac{\mu\!-\!\mu_c}{T},\frac{h\!-\!h_c}{T}\!\!\rp
\label{eq:scaling-conservedcharges}
\ee
where $x=1+d/z$ for $\mc O = n_{\sigma}$ and $x=d/z$ for $\mc O=\kappa$.
Note that here we include two relevant couplings, $h$ and $\mu$, and these are the only two relevant static couplings for the DFG universality class we consider below. Close to critical point one can Taylor expand the regular pieces $n_{\sigma,r}$ and $\kappa_r$ and keep only the constant terms, which we denote $n_{\sigma}^{(0)}$ and $\kappa^{(0)}$. For the MI/metal crossover, the constant term for the regular piece for density and compressibility are intuitively expected to be $n^{(0)}=1$ and  $\kappa^{(0)}=0$, and the calculations below confirm this.

This scaling structure also applies to some more general transitions than symmetry breaking bosonic ones, including the DFG transition of interest here~\cite{sachdev_quantum_2001}.  This theory is defined by the Hamiltonian
\be
H_{DFG} &=& \sum_{\v{k},\sigma} \lp \frac{\hbar^2 k^2}{2m^*} - \mu^*\rp n_{\v{k}\sigma}+ \sum_{\v{k}} h^* \lp n_{\v{k}\uparrow}-n_{\v{k}\downarrow}\rp  \nonumber \\
    &&\hspace{0.1in}{}+ \frac{g}{2}\sum_{\v{p},\v{k},\v{q}} c^\dagger_{\v{k},\uparrow}c^\dagger_{\v{p},\downarrow} c_{\v{p}-\v{q},\downarrow}^{\phantom{\dagger}} c_{\v{k}+\v{q},\uparrow}^{\phantom{\dagger}} \label{eq:DFG-Ham}
\ee
where $m^*$ is the effective mass, $g$ the effective interaction, $\mu^*$ the effective chemical potential,  $h^*$ the effective magnetic field,  $c^{(\dagger)}_{\v{k}\sigma}$
are fermionic annihilation (creation) operators for momentum $\v{k}$ and spin $\sigma$, and $\rho_{\v{k}\sigma}\equiv c^\dagger_{\v{k}\sigma}c^{\phantom{\dagger}}_{\v{k}\sigma}$.  For the present case, it can be shown that $\mu^*=\mu-\mu_c$ and $h^* = h-h_c$.
The interaction term is the only relevant coupling in $d<2$ and is marginal in $d=2$.  We stress that this is a (candidate) effective model only near the quantum critical crossover, and there is no simple relationship between the renormalized parameters $g$ and $m^*$ and the microscopic parameters $U$, $\mu$, $h$, and $t$. Our DQMC calculations are for the Hubbard model, Eq.~\eqref{eq:Hubb}.

Despite the fairly simple nature of the Hamiltonian in Eq.~\eqref{eq:DFG-Ham}, it   displays a quantum phase transition with much of the phenomenology of general quantum critical behavior.  The $T=0$ phase transition is tuned by $\mu$ and occurs at $\mu_c=0$.  For $\mu^*<0$, the zero temperature system is a vacuum of no particles, and at temperatures $T\ll \mu^*$ consists of a dilute classical gas.  For $\mu^*>0$, the zero-temperature system is a Fermi liquid, and low finite temperatures $T\ll|\mu^*|$ add dilute quasiparticle excitations.  At finite temperatures satisfying $T\gsim |\mu^*|$, an intervening ``quantum critical" region occurs, and the spatial separation between excitations is comparable to their thermal de Broglie wavelength.

\begin{figure}[t]
\setlength{\unitlength}{1.0in}
\includegraphics[width=2.55in,angle=0]{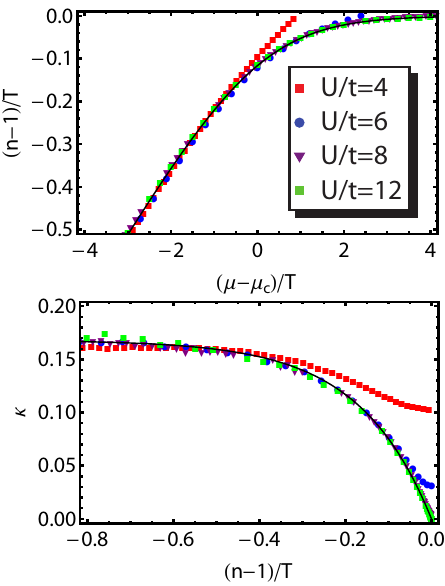}
\vspace{-0.2in}
\caption{ (Color online) Universal $\mu$-dependence of the density and compressibility for several tunneling rates and interaction strengths ($t/U$'s).
Data are for $T=t/3$, except for $U/t=12$ for which $T=t/2.5$.
We observe excellent universal collapse for all data at $U/t=6,8,12$
and also for all but very low hole densities at $U/t=4$. 
Top: $(n-1)\xi/T$ vs. $(\mu-\mu_c)/T$ for $h=0$, where $\xi=m^*/(2\pi \hbar^2)$.
Bottom: $\kappa\xi$ vs. $(n-1)\xi/T$.  We omit $\xi$'s in plot labels throughout, as is conventional.
We fit the non-universal scaling factor $\xi$
to  obtain collapse with the dilute Fermi gas critical theory prediction, Eq.~\eqref{eq:DFG-noninteracting}.
 \label{fig:univ-collapse}}
\end{figure}

If we ignore the marginal interaction term, the DFG  observables are simply those of a non-interacting gas:
\be
n_{\sigma}(\mu,h,T) &=& \frac{m^* T}{2\pi\hbar^2} \log \lp 1+e^{\beta (\mu+ \sigma h)}\rp \nonumber \\
\kappa(\mu,h,T) &=& \frac{m^* }{2\pi\hbar^2}\lp\frac{1}{e^{\beta(\mu-h)}+1}+\frac{1}{e^{\beta(\mu+h)}+1}\rp \label{eq:DFG-noninteracting}
\ee
with $\beta=1/T$, and $\sigma$ in the exponential takes values $\{+1,-1\}$ for $\sigma= \{\uparrow,\downarrow\}$.  We expect the interactions to give logarithmic corrections in $d=2$, but we find that these are negligible in the regime considered herein.

Because the density is a derivative of the free energy with respect to $\mu$, a relevant coupling, it will have a singular, universal contribution.  Indeed, the only universal static observables in the DFG theory are the density and magnetization (and their derivatives with respect to $\mu$ and $h$).  Other observables, for example the double occupancy or nearest neighbor spin correlations, are derivatives of the free energy with respect to irrelevant couplings and thus have no singular, universal contribution.  Consequently, these other observables are non-universal.


\textit{Results}---We compute several observables, enumerated below, with DQMC for a $L\times L$ lattice with $L=10$, discretizing imaginary time evolution into Trotter steps of size $\delta \tau = 1/(12t)$, running $1000$ Monte Carlo equilibration steps, followed by $100,000$ steps for statistical sampling, for each value of $\mu$ and $h$.  The one exception was the $t/U=1/12$, $T/t=1/3$ calculation, for which $500,000$ statistical sampling steps were necessary to get sufficiently accurate results.  Typical statistical error bars for the density, not shown, are less than a tenth of a percent, smaller than the point size.  We have checked convergence in each of these parameters.  We found that the results were unchanged roughly within statistical error bars going from $L=8$ to $L=16$  for the $t/U=1/4$ case where we expect the finite-size effects to be most important. We similarly found no change within statistical error bars from decreasing the Trotter step to $1/(16t)$.

To determine $\mu_c$, $n_\sigma^{(0)}$, and $\kappa^{(0)}$, we plot $n/T$ versus $\mu$ (for the $h=0$ calculations).
Eqs.~\eqref{eq:scaling-conservedcharges} imply that for observables in the scaling region, the resulting curves should cross at $\mu=\mu_c$ with the crossing value giving $\mc O^{(0)}$.From this, we find that $n^{(0)}=1$,  $\kappa^{(0)}=0$, and $\mu_c/t=\{0,-0.83,-1.2,-3\} $ for $U/t=\{4,6,8,12 \}$.  We also note that plotting $\kappa$ versus $n/T $ gives a universal curve, so one can circumvent determining $\mu_c$.
Note that the $U/t=4$ value of $\mu_c$ is likely non-zero, but is zero within our error bars. The $\mu_c/t$ used in the scaling analysis below are slightly different than these, namely $\mu_c/t=\{-0.3,-0.8,-1.5,-3.2\}$. These are in rough agreement with the crossing points, but shifted by $ \delta (\mu_c/t) \approx0.2$. These are found by adjusting to get best agreement with DFG theory.  The difference between the two methods is likely due to the fact the crossing is 
influenced by data from non-universal temperatures.
 It is worth comparing these values of $\mu_c/t$ to expectations in the small- and large-$t/U$ limits.  Note the trend of decreasing $\mu_c$ for increasing $t$: as expected, the MI regime shrinks as $t$ increases.
For small $t/U$, we can compare this to naive expectations from a simple model of Hubbard bands with width $8t$ that $\mu_c/t = -U/(2t) +4$, which gives $\mu_c/t = -2$ for our smallest $t/U=1/12$, in rough agreement with our observations.  For large $t/U$, the RPA treatment of the antiferromagnet gives $\mu_c \approx -t e^{-2\pi \sqrt{t/U}}$.  For our smallest $t/U=1/4$, this is  $\mu_c/t  = -0.04 $, also in rough agreement with our observations, although this data is far from the large $t/U$ limit.

\begin{figure}[hbtp]
\setlength{\unitlength}{1.0in}
\includegraphics[width=3.4in,angle=0]{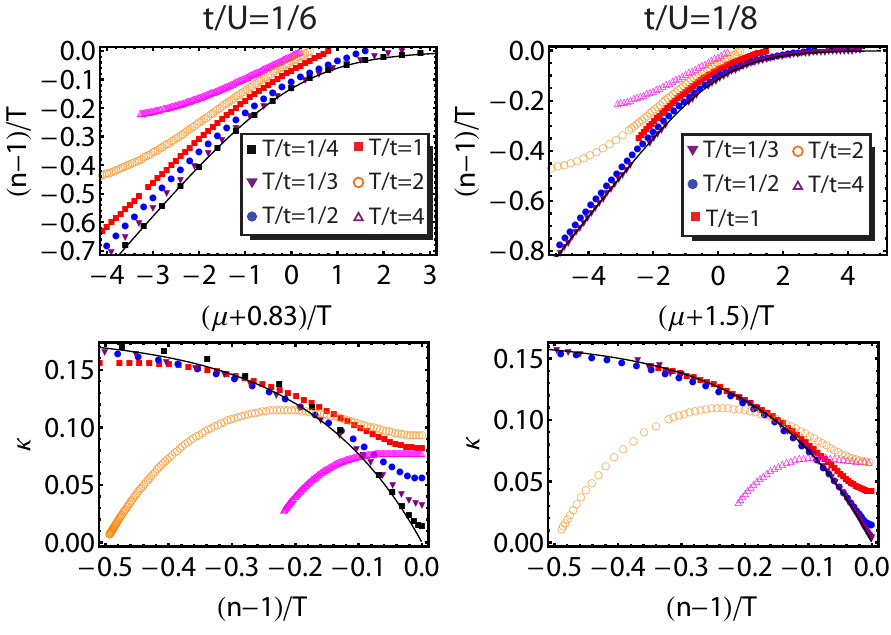}
\vspace{-0.35in}
\caption{(Color online) Universal $\mu$-dependence
of the densities and compressibility.
 Universality breaks down outside of $t^2/U\lsim T \lsim t$ window.
Left column: $U/t=6$, Right column: $U/t=8$.
 Although the collapse for $U/t=6$ is fairly good at the lowest temperatures, the collapse for $U/t=8$ is better for a wider range of temperatures, as expected since the universal temperature window is larger.
\label{fig:univ-and-non-univ-regimes}}
\end{figure}


Figure~\ref{fig:univ-collapse} shows that the density and compressibility are universal functions of $\mu$ for temperatures $T=t/3$ (all but $t/U=1/12$) or $t/2.5$ ($t/U=1/12$, because it is difficult to get accurate compressibilities at lower temperatures).  Each panel shows shows the appropriately rescaled observables for  $t/U=1/4,1/6,1/8,$ and $1/12$.  The collapse  indicates $z=2$ scaling behavior for several $t/U$'s at $t/T = 1/3$. Furthermore, Fig.~\ref{fig:univ-collapse} shows quite remarkably that the non-interacting DFG expressions, Eq.~\eqref{eq:DFG-noninteracting}, quantitatively describe the scaling functions of the Hubbard model in the strongly correlated regime near the MI/metal crossover.  Although in $d=2$ we expect logarithmic corrections to the non-interacting scaling functions,  these are unobservably small.

Fig.~\ref{fig:univ-and-non-univ-regimes} gives further evidence for this by showing collapse of data at multiple temperatures for $t/U=1/8$ and $t/U=1/6$, and we find that the DFG quantitatively describes the $\mu$-dependence
 for temperatures satisfying $t^2/U \lsim T \lsim t$, a fairly natural result.  The $T=t/3$ results  in Fig.~\ref{fig:univ-collapse} are also in this temperature window.
We also have checked that, in contrast to the plots of rescaled variables, e.g. $(n-1)/T$ vs $(\mu-\mu_c)/T$, plotting variables without rescaling or plotting rescaled variables corresponding to other $z$'s, e.g. $z=1$ or $z=4$, does \textit{not} give good collapse (see Supplementary Information).

The data collapse observed for $t^2/U \lsim T \lsim t$ breaks down for both lower and higher temperatures.  At higher temperatures this is natural because the excitations begin to probe energy scales beyond the bottom of the excitation band, where the dispersion is well described by $\epsilon_k = \hbar^2 k^2/(2m^*)$, and begin to see  the microscopic band structure.  At lower temperatures, the spins are no longer independent, and develop magnetic correlations.  These correlations at high temperatures depend on microscopic details governing superexchange physics.  Only at much lower temperatures, $T\ll t^2/U$ will universal behavior recover, and it will likely be more exotic, for example one of the scenarios referenced in the introduction.

To compare scaling functions obtained with DQMC to the DFG theory's, one requires a single non-universal scaling factor, the DFG effective mass in Eq.~\eqref{eq:DFG-noninteracting}.  From this,  we find the ratios of $m^*$'s for $t/U=1/4,1/6,1/8,1/12$ are $1:1.7:2.4:4.1$.  The $m^*$ increases as $t/U$ decreases, roughly as $1/t$ for small $t$.

Figures~\ref{fig:univ-collapse} and~\ref{fig:univ-and-non-univ-regimes} illustrate that universal scaling is best for small values of $t/U$, becoming quantitative over the full visible range only for $t/U \lsim 1/8$.  This is natural, since only for small $t/U$ is there a large separation between the $t^2/U$ and $t$ temperature scales that must hold in order for the temperature to satisfy $t^2/U \ll T \ll t$.  

In the DFG universality class, $n_\uparrow$, $n_\downarrow$, and $\kappa$ and their derivatives are the only universal static observables.  Figs.~\ref{fig:univ-collapse} and~\ref{fig:univ-and-non-univ-regimes} confirmed their universality.
We also confirmed that other observables are non-universal: the nearest neighbor spin correlations, doubly occupied sites, and kinetic energy (see Supplementary Information).

In contrast to naive expectations, we find that the DFG fails to capture the $h$-dependence of the spin densities.
The reason is that in addition to the itinerant charge carriers governed by the DFG, there is a Mott background with spin degrees of freedom.  For $t/U\ll 1$, this background is described by a high temperature $T\gg t^2/U$ Heisenberg model, which reduces to a  single site problem.  The spin densities from this background are $n_\uparrow = 1/(1+e^{2\beta h})$ and $n_\downarrow = 1/(1+e^{-2\beta h})$.  At $T=0$, these are singular (step functions) at  $h=0$, indicating a non-analytic contribution to the spin densities that is not captured by the dilute Fermi gas theory.  Thus at $\mu=\mu_c$ for $h=0$, there are coinciding singularities from this ``background" and DFG contributions.  Rather than simply summing, it appears they may combine to give new universal behavior, with $z\approx 4$.
This is elaborated  in the Supplementary Information.


\textit{Summary.}---We have computed the density, compressibility, fraction of doubly occupied sites, kinetic energy, and nearest neighbor spin correlations of the two-dimensional square lattice Hubbard model with determinantal quantum Monte Carlo near the Mott/metal crossover. Resolving a long discussion in the literature, we find that the behavior is quantitatively described by the dilute Fermi gas universality class.  For the observables that this class predicts to be universal --- spin densities and compressibilities --- we find the dilute Fermi gas theory quantitatively gives the scaling functions at $h=0$ for a range of $t/U\lsim 1/6$ and temperatures $t^2/U\lsim T \lsim t$.  We also confirm that there is no universality for observables the dilute Fermi gas theory predicts to be non-universal.
For $h\ne 0$, the singular contribution from the non-itinerant spin degrees of freedom invalidates the dilute Fermi gas theory, and interaction of localized spins with itinerant carriers may give new universal physics.

\textit{Acknowledgements.}---KH and AMR acknowledge support from grants from the NSF (PFC and PIF-0904017), the AFOSR, and a grant from the ARO with funding from the DARPA-OLE. RTS was supported under ARO Award W911NF0710576 with funds from the DARPA OLE Program.


\clearpage

\title{Supplementary information for: Universality class of quantum criticality in the two-dimensional Hubbard model at high-temperature ($t^2/U\ll T\ll t$) }
\author{Kaden R.~A. Hazzard} \email{kaden.hazzard@colorado.edu}
\affiliation{JILA and Department of Physics, University of Colorado, Boulder, and NIST, Boulder, Colorado 80309-0440, USA}
\author{Ana Maria Rey}
\affiliation{JILA and Department of Physics, University of Colorado, Boulder, and NIST, Boulder, Colorado 80309-0440, USA}
\author{Richard T. Scalettar}
\affiliation{Physics Department, University of California, Davis, California 95616, USA}

\maketitle

\section{Comparison of rescaling for $z=1$, $z=2$, and $z=4$}

\begin{figure}[h]
\setlength{\unitlength}{1.0in}
\includegraphics[width=2.6in,angle=0]{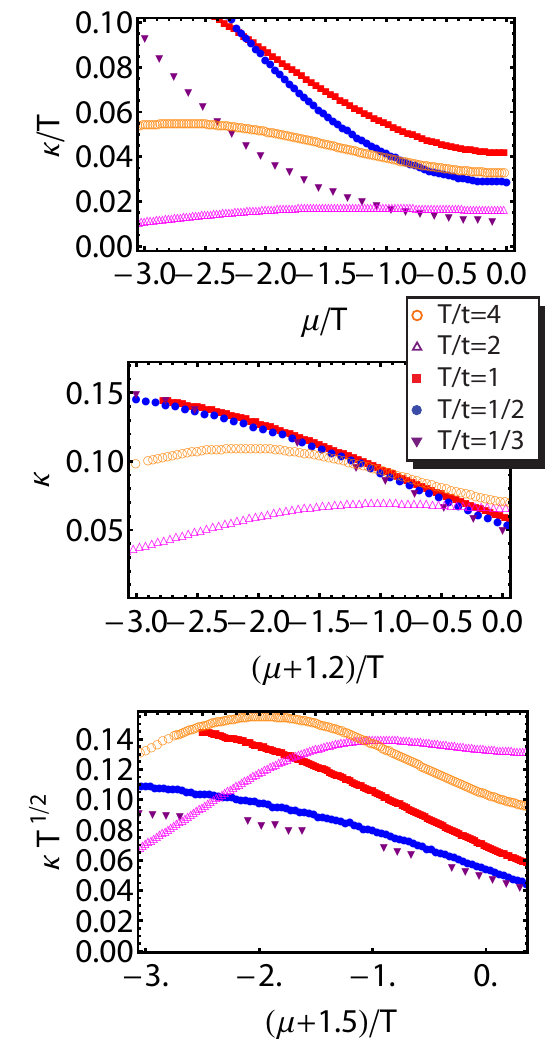}
\caption{Rescaled compressibility ($\kappa/T^{d/z-1}$) versus rescaled chemical potential $(\mu-\mu_c)/T$ for $U/t=8$ and dynamic exponent $z=1,2,4$, top to bottom.
 The $z=2$ rescaling obtains good collapse while the $z=1$ and $z=4$ both fail  to give any significant collapse. The other values of $U/t$ we considered show similar results. \label{fig:univ-and-non-univ-compare}}
\end{figure}

Figure~\ref{fig:univ-and-non-univ-compare} illustrates that the $z=2$ rescalings offer a good universal collapse of the data, while other candidate $z$'s fail.  We show rescalings associated with $z=1,2,$ and $4$, and only the $z=2$ case provides an adequate description of the data.  We determined $\mu_c$ from the best crossing point of $n/T^{d/z}$, but for $z=1$ and $z=4$, there were no good crossings ($\mu_c=0$ and $\mu_c=-1.5$, respectively, were the best or at least as good as any others).  Note that in the main text, the presented data used $\mu_c$ determined to give good collapse of the low-temperature data to the dilute Fermi gas theory, and the resulting value of $\mu_c$ was slightly different than with the crossing method (e.g., $\mu_c=-1.5$ instead of $\mu_c=-1.2$ for the $z=2$ rescaling of the $U/t=8$ data).  Since we have no natural theory to use for comparison in the $z=1$ and $z=4$ cases, determining $\mu_c$ by the crossing points was more fair for the comparison of Fig.~\ref{fig:univ-and-non-univ-compare}.

\section{Failure of non-universal observables to rescale}

\begin{figure}[h]
\setlength{\unitlength}{1.0in}
\includegraphics[width=2.6in,angle=0]{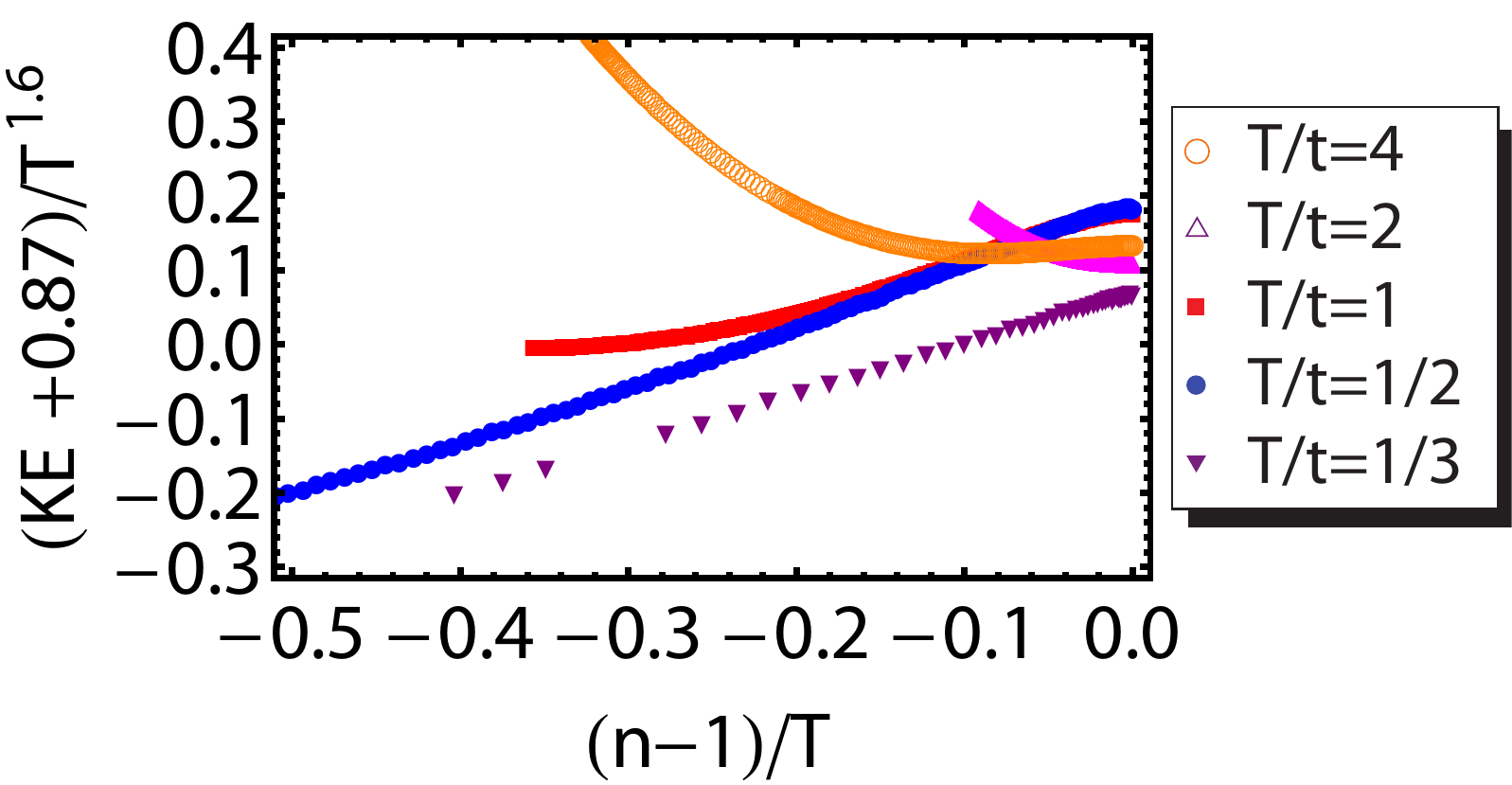}
\caption{Rescaled kinetic energy $(K+0.87)/T^{1.6}$ versus rescaled density $(n-1)/T$, for $t/U=1/8$.  We have adjusted  the offset (0.87) and the exponent (1.6) to find the best collapse for the two lowest temperature curves, but collapse is still poor. \label{fig:kedoesntrescale}}
\end{figure}

As discussed in the main text, besides the $\mu$- and $h$-derivatives of the free energy, for example the densities and compressibilities, we expect no other static universal observables.  We have confirmed that other observables indeed do not collapse: the kinetic energy $K$, nearest neighbor spin correlator $X$, and double occupation probability $D$. This remains the case, even if we allow the non-universal constant, $D(T=0,\mu=\mu_c)$, and the scaling exponent to take arbitrary values.  Fig.~\ref{fig:kedoesntrescale} illustrates the best collapse we could obtain for the kinetic energy at $t/U=1/8$, found by matching the small hole density behavior of the two lowest temperature curves.  As expected, the collapse is very poor.  These results are typical.

\section{$h$-dependence of observables disagrees with DFG theory}

\begin{figure}
\setlength{\unitlength}{1.0in}
\includegraphics[width=2.7in,angle=0]{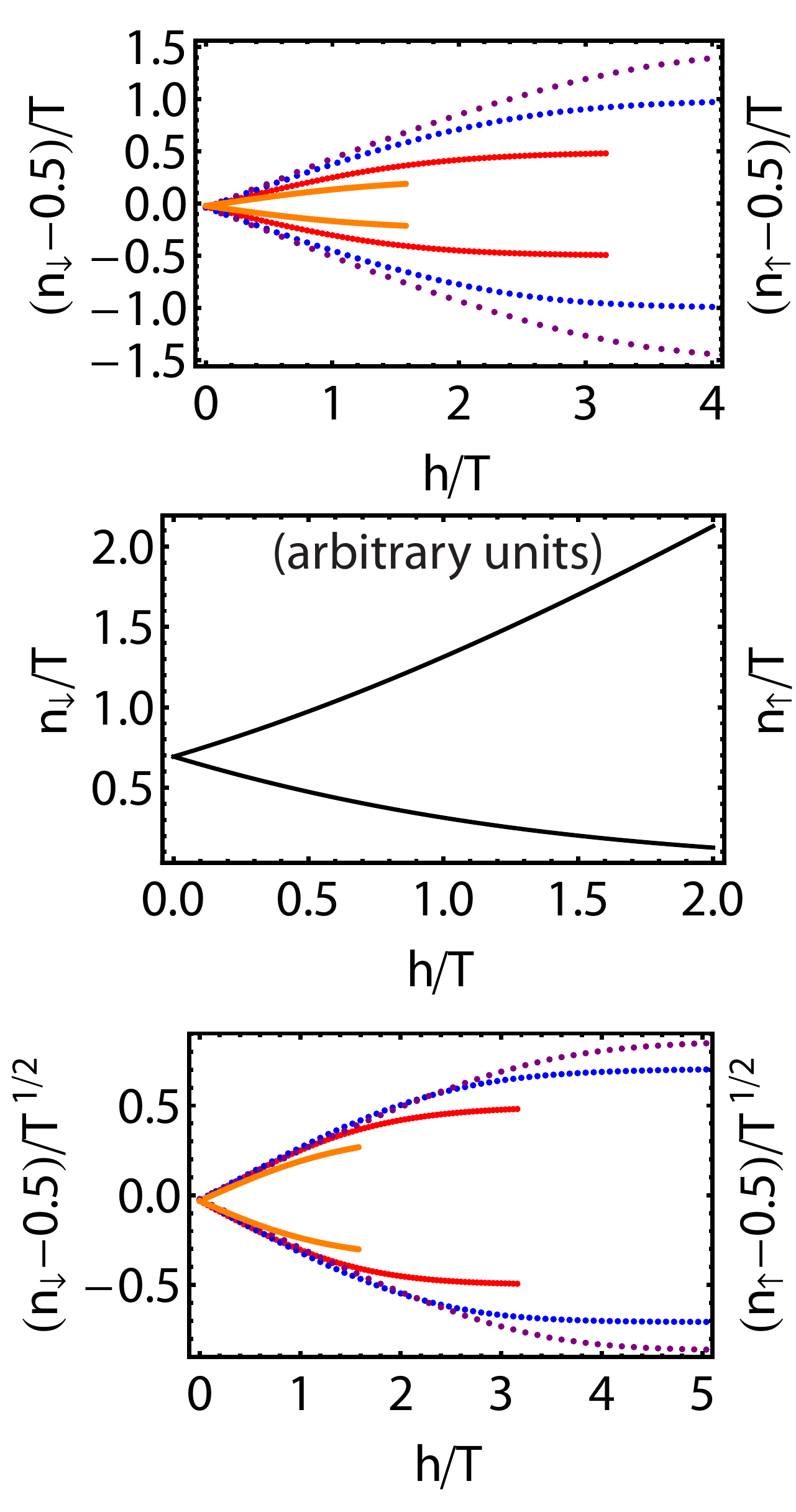}
\caption{
Top: determinantal quantum Monte Carlo data for spin-down density  (top curves) and spin-up density (bottom curves) rescaled at $U/t=8$, assuming a dynamical critical exponent $z=2$. Going from ``inner curves" to ``outer curves," the temperatures are temperatures $T/t=2,1,1/2,1/3$.  No scaling collapse is observed.
Middle:  dilute Fermi gas theory predictions for comparison.
Bottom: same as top, but results rescaled according to $z=4$.
\label{fig:hsweep-results}}
\end{figure}

We have noted that $\mu$- and $h$-are (the only) relevant tuning parameters within the dilute Fermi gas (DFG) theory.  Thus we might expect the DFG to govern the universal $h$-dependence of observables.  However, we have computed the spin-up and spin-down densities of the Hubbard model as a function of $h$ for $\mu=\mu_c$ and $t/U=1/8$ and find that the DFG theory does not adequately describe these observables.  More strongly, no scaling collapse is obtained for $z=2$, ruling out not only the DFG universality class, but any $z=2$ theory.  Fig.~\ref{fig:hsweep-results} (top panel) demonstrates the lack of $z=2$ scaling, and the middle panel shows the DFG theory curves for comparison.

To understand this lack of collapse, we reconsider the physical picture leading to the application of the dilute Fermi gas theory. We also note that no formal justification has to this point been available, so that we must rely on our physical intuition.  The basic picture was of itinerant dopants governed by the DFG theory coupled to an essentially inert Mott background.  The Mott background's only effect was to renormalize the DFG parameters and provide a non-universal, non-singular contribution to the observables, which was irrelevant near the quantum critical point.

However, when a magnetic field is applied, treating the Mott background as inert is dangerous. To see this, consider the case where there are \textit{no} itinerant charge carriers and thus the DFG plays no role.  The Mott background may be described by a Heisenberg model, and since we are in the high temperature limit $T\gg t^2/U$, this reduces to a single site problem.  In this case, the spin densities are $n_\uparrow=e^{-\beta h}/\lp e^{-\beta h}+e^{\beta h}\rp$ and $n_\downarrow=e^{\beta h}/\lp e^{-\beta h}+e^{\beta h}\rp$.  We see these are singular (step functions) at $h=0$ and $T=0$, so that the ``Mott background" gives a singular contribution here, rather than a regular contribution as in the $h=0$ case, tuning $\mu$.  This indicates that the singular contribution will include both a universal contribution from the DFG theory and from the ``background."  The extent to and way in which these two pieces of physics couple at $h\ne 0$ is a complicated issue that we leave to future work.

Interestingly, we do observe rather good collapse for $z=4$, reminiscent of the $z=4$ exponent found at $T=0$ where magnetic physics, including superexchange, is certainly relevant~\cite{imada_metal-insulator_1998}.  However, we emphasize we are well above the superexchange temperature.  It would be intriguing to understand this result better, and in particular if it indicates a new universality class governing the $t^2/U\ll T \ll t$ Hubbard model.  Presumably such a universality class would consist of free spins coupled to a DFG in some way, although to be consistent with our results in the main text it must reduce to the DFG universality class when $h=0$.


\end{document}